\documentstyle[editedvolume,numreferences,epsfig]{crckapb} 

\def\FigSizeBraneBreaking{0.7\textwidth}
\def\FigSizeDThreeConfig{0.6\textwidth}
\def\FigSizeElecField{0.9\textwidth}
\def\FigSizeFiveBraneConfig{0.8\textwidth}
\def\FigSizeGeneralCase{0.8\textwidth}
\def\FigSizeNonLagrangian{0.4\textwidth}
\def\FigSizeSimpleBrane{0.5\textwidth}
\def\FigSizeSimpleQuiver{0.5\textwidth}
\def\FigSizeTransition{0.9\textwidth}
\def\FigSizeUkGeneralFlavor{0.8\textwidth}

\newcommand{\beq}{\begin{equation}}
\newcommand{\eeq}{\end{equation}}
\newcommand{\beqar}{\begin{eqnarray}}
\newcommand{\eeqar}{\end{eqnarray}}
\newcommand {\nono} {\nonumber \\} 

\newcommand {\bear} [1] {\begin {array} {#1}}
\newcommand {\ear} {\end {array}}

\newcommand {\eqr} [1] 
	{(eq. \ref {eq:#1})}

\newcommand {\beqarn} {\begin{eqnarray*}}
\newcommand {\eeqarn} {\end{eqnarray*}}

\newcommand	{\abs}	[1] {{\left| #1 \right|}}

\newcommand {\mrm} [1] {\mathrm {#1}}

\newcommand {\myref} [1]	%
	{%
	\begin{thebibliography} {99}	%
			{#1}	%
	\end {thebibliography}}

\def\emph#1{{\em #1}}
\def\mathbf#1{{\bf #1}}
\def\mathrm#1{{\rm #1}}
\def\mathit#1{{\it #1}}

\catcode`\@=11
\@addtoreset{footnote}{section}
\catcode`\@=12

\newcounter	{exinsert}	[subsection]

\renewcommand {\theexinsert}	%	
{	\thesubsection.\arabic{equation}}

\renewcommand {\theequation} {\thesection.\arabic{equation}}

\catcode`\@=11
\@addtoreset{equation}{section}
\catcode`\@=12

\newenvironment {exinsert} [1]	%
{	%begindef	
	\begin {quotation}	%
	\refstepcounter {equation}	%
	{\bf {#1}} \theequation. \,\,	%
}	
{	%enddef	%
	\end {quotation}	
}

\newcommand {\bprop} 
{\begin {exinsert} {Proposition}
}
\newcommand {\eprop} {\end {exinsert} }

\newcommand {\bexe} {\begin {exinsert} {Exercise}}
\newcommand {\eexe} {\end {exinsert} }

\newcommand {\bexa} {\begin {exinsert} {Example}}
\newcommand {\eexa} {\end {exinsert} }

\newcommand {\literal} [1] {{ \mbox {$\mrm {#1}$}}}

\newcommand {\grad} {\nabla}

\newcommand {\ncmd} [2] {\newcommand {#1} {#2}}

\ncmd {\vgrad} {{\vec \grad}}
\newcommand {\dod} [1] {\frac{\partial}{\partial #1}}

\ncmd {\Mv} {{\cal M}_V}
\ncmd {\Mh} {{\cal M}_H}

\ncmd {\adj} {\literal {adj}}	%	Adjoint
\ncmd {\Mp} {M_\literal {planck}}	%	Planck mass
\def\FI{Fayet-Iliopoulos }

\def\kahler{K\"ahler }

\ncmd {\x} {$\times$}

\newcommand {\myfigure} [4]
  {\begin{figure} [htb]
   \begin{center}
	\epsfig{figure=#3, width=#1}
   \end{center}
   \caption {#4}	\label {fig:#2}
  \end{figure}}

\newcommand {\putfig} [3] {\myfigure {#2} {#1} {#1.eps} {#3}}

\newcommand {\figr} [1]	{figure \ref {fig:#1}}

\newcommand {\btab} [3]
  {\begin{table} [htb]
   \begin{center}
   \caption {#2}	\label {tab:#1}
   \begin {tabular} {#3}
   \hline}

\ncmd {\etab}	{   \hline
   \end {tabular}
   \end{center}
  \end{table}}

\newcommand {\tabr} [1]	{table \ref {tab:#1}}

\def\jou#1,#2,#3,{{\sl #1\/ }{\bf #2} (19#3)\ }
\def\ibid#1,#2,{{\sl ibid.\/ }{\bf #1} (19#2)\ }
\def\am#1,#2,{{\sl Acta. Math.\/ } {\bf #1} (19#2)\ }
\def\annp#1,#2,{{\sl Ann.\ Physics\/ }{\bf #1} (19#2)\ }
\def\cmp#1,#2,{{\sl Comm.\ Math.\ Phys.\/ }{\bf #1} (19#2)\ }
\def\cqg#1,#2,{{\sl Class.\ Quantum Grav.\/ }{\bf #1} (19#2)\ }
\def\dm#1,#2,{{\sl Duke\ Math.\ J.\/ }{\bf #1} (19#2)\ }
\def\ijmpa#1,#2,{{\sl Int.\ J.\ Mod.\ Phys.\/ }{\bf A#1} (19#2)\ }
\def\invm#1,#2,{{\sl Invent.\ Math.\/ }{\bf #1} (19#2)\ }
\def\jhep#1,#2,{{\sl JHEP\/ }{\bf #1} (19#2)\ }
\def\jmp#1,#2,{{\sl J.\ Geom.\ Phys.\/ }{\bf #1} (19#2)\ }
\def\jams#1,#2,{{\sl J.\ Diff.\ Geom.\/ }{\bf #1} (19#2)\ }
\def\jdg#1,#2,{{\sl J.\ Amer.\ Math.\ Soc.\/ }{\bf #1} (19#2)\ }
\def\jpa#1,#2,{{\sl J.\ Phys.\ A.\/ }{\bf #1} (19#2)\ }
\def\grg#1,#2,{{\sl Gen.\ Rel.\ Grav.\/ }{\bf #1} (19#2)\ }
\def\mpla#1,#2,{{\sl Mod.\ Phys.\ Lett.\/ }{\bf A#1} (19#2)\ }
\def\ma#1,#2,{{\sl Math.\ Ann.\/ } {\bf #1} (19#2)\ }
\def\nc#1,#2,{{\sl Nuovo Cim.\/ }{\bf #1} (19#2)\ }
\def\npb#1,#2,{{\sl Nucl.\ Phys.\/ }{\bf B#1} (19#2)\ }
\def\plb#1,#2,{{\sl Phys.\ Lett.\/ }{\bf #1B} (19#2)\ }
\def\pla#1,#2,{{\sl Phys.\ Lett.\/ }{\bf #1A} (19#2)\ }
\def\pnas#1,#2,{{\sl Proc.Nat.Acad.Sci.\/ }{\bf #1} (19#2)\ }
\def\prev#1,#2,{{\sl Phys.\ Rev.\/ }{\bf #1} (19#2)\ }
\def\prd#1,#2,{{\sl Phys.\ Rev.\/ }{\bf D#1} (19#2)\ }
\def\prl#1,#2,{{\sl Phys.\ Rev.\ Lett.\/ }{\bf #1} (19#2)\ }
\def\prpt#1,#2,{{\sl Phys.\ Rept.\/ }{\bf #1C} (19#2)\ }
\def\ptp#1,#2,{{\sl Prog.\ Theor.\ Phys.\/ }{\bf #1} (19#2)\ }
\def\ptpsup#1,#2,{{\sl Prog.\ Theor.\ Phys.\/ Suppl.\/ }{\bf #1} (19#2)\ }
\def\rmp#1,#2,{{\sl Rev.\ Mod.\ Phys.\/ }{\bf #1} (19#2)\ }
\def\sm#1,#2,{{\sl Selec. Math.\/ }{\bf #1} (19#2)\ }
\def\yadfiz#1,#2,#3[#4,#5]{{\sl Yad.\ Fiz.\/ }{\bf #1} (19#2) #3%
\ [{\sl Sov.\ J.\ Nucl.\ Phys.\/ }{\bf #4} (19#2) #5]}
\def\zh#1,#2,#3[#4,#5]{{\sl Zh.\ Exp.\ Theor.\ Fiz.\/ }{\bf #1} (19#2) #3%
\ [{\sl Sov.\ Phys.\ JETP\/ }{\bf #4} (19#2) #5]}

\def\argyres	{P.~Argyres}

\def\boer	{J.~de~Boer}

\def\douglas	{M.~R.~Douglas}

\def\green	{M.~B.~Green}

\def\hori	{K.~Hori}
\def\intriligator {K.~Intriligator}
\def\moore	{G.~Moore}

\def\oz		{Y.~Oz}
\def\ooguri	{H.~Ooguri}
\def\plesser	{M.~R.~Plesser}

\def\schwarz	{J.~H.~Schwarz}
\def\seiberg	{N.~Seiberg}

\def\strominger	{A.~Strominger}
\def\vafa	{C.~Vafa}

\def\witten	{E.~Witten}

\ncmd {\va} {\vec{a}}
\ncmd {\vr}{\vec{r}}
\ncmd {\vm} {\vec{m}}
\ncmd {\vomega} {\vec{\omega}}
\ncmd {\vzeta} {\vec{\zeta}}

\ncmd {\MvA} {{{\cal M^A}_V}}
\ncmd {\MvB} {{{\cal M^B}_V}}
\ncmd {\MhA} {{{\cal M^A}_H}}
\ncmd {\MhB} {{{\cal M^B}_H}}

\begin{opening}

\title{Mirrors and Phases of N=4 in D=3}
\subtitle{\hfill \rm \small UCB-PTH-97/64, LBNL-41163, hep-th/9712184}

\author{Zheng Yin}
\institute{Department of Physics, 
University of California\\
Berkeley, CA 94720-7300, U.S.A.\\
{\sl and}\\
Mail Stop 50A-5101 (zyin@thsrv.lbl.gov)\\
Lawrence Berkeley National Laboratory \\
1 Cyclotron Road, 
Berkeley, CA 94720, U.S.A.}

\end{opening}

\runningtitle{Mirrors and Phases}

\begin{document}

\begin{abstract}
We review brane engineering of mirror pairs of 3d N=4 theories.  
It reveals aspects of 3d physics not known from 
previous field theoretic studies: novel QFT's without 
Lagrangian description and transitions to them from  
conventional QFT's.
\end{abstract}

\section{Introduction}

	3d mirror symmetry \index{mirror symmetry} for was first 
proposed in \cite {is} as a duality between 
certain pairs of generally different 3d N=4 theories 
at the infrared limit.  Infinite 
sequences of new mirror pairs and strong field theoretic 
evidence for them were found in \cite {bhoo}.  It is a 
nonperturbative duality that in particular equates certain quantities 
receiving large quantum corrections with some that are determined 
entirely classically.  Naturally one asks whether this can be 
the consequence of some string dualities.  There are several 
different approaches \cite {is,bhoo,pz,hw,bhooy,hov,kmv:mirror-four-d}, 
each of which has it own advantage and is related to the others by some 
sequences of dualities.  In this talk I will review one that is 
particularly intuitive \cite {hw,bhooy}.  The global R-symmetry of 
N=4 theories appears as geometric rotations; R-symmetry breaking part 
of the moduli space of vacua and the 
parameters of the field theory are realized as the moduli space of 
D-branes configurations; mirror symmetry itself is implemented by 
the S duality of type IIB string theory \cite {hw}.  This construction 
allows us to engineer a large class of theories  
and find their mirror duals \cite {bhooy}.  Since then it has been 
generalized to 3d N=2 theories \cite {bhoy,ahiss}.  Similar ideas of 
constructing field theories in 4d, reviewed in Eguchi's lecture 
at this school \cite {eguchi:cargese}, have also been very fruitful\footnote 
	{See, for example, 
	\cite {egk,w:four-d,hoo:strong,w:qcd,bhoo:kahler}.}.
In this talk, I will focus on 3d N=4 theories.  After reviewing 
the rules for ``model building'' via ``brane engineering,'' I will show 
how the mirror pairs emerge from this prescription.  As an unexpected 
reward we can predict an infinite number of 3d field 
theories without conventional Lagrangian descriptions.  Some of 
them are dual to ordinary Lagrangian theories via mirror symmetry, 
but the rest are not; yet they can be smoothly connected in the moduli 
space of brane configurations.  

\section{What is 3d Mirror Symmetry?}

The structure of N=4 supersymmetric gauge theories in three dimensions 
can be easily obtained by  
dimensionally reducing the minimally supersymmetric 6d Yang-Mills 
Lagrangian \cite {sw:three-d}.  The global R-symmetry \index{R-symmetry} 
of the 3d theory is $SU(2)_{345} \times SU(2)_{789}$.  The 
reason for choosing these subscripts will become clear 
in the brane realization 
described later.  The field content consists of vector 
multiplets and hypermultiplets.
For each vector multiplet associated with a $U(1)$ factor of the gauge group 
there are 3 real \FI  parameters \index{Fayet-Iliopoulos parameter}, 
$\vzeta$, which can be thought as coming 
from the VEV's of a background hypermultiplet.  
For each hypermultiplet, there are 3 real mass parameters, $\vm$.  They are 
the VEV's of a background vector multiplet.  There are also 
gauge coupling 
constants, which also come from background vector multiplets.  
The transformation 
properties of the parameters and VEV's  under R-symmetry  
are summarized in \tabr {r-charge}.  

	Note that usually the scalars in a hypermultiplet 
are written as a doublet under $SU(2)_{789}$.  However, it is 
convenient, by a change of variables, to rearrange them into a singlet $b$ 
and a triplet $\vec r$.  
On the other hand, an interesting feature 
peculiar to three dimensions is that a vector 
potential is dual to a scalar by the usual electro-magnetic duality 
\index{electro-magnetic duality}.  
Of course this duality transformation can be precisely 
formulated only for a free 
$U(1)$ gauge fields, but this is what is available for the low 
energy effective theory at generic points of the vector multiplet branch 
of moduli space.  Therefore 
it is meaningful to include the dualized scalar in considering the moduli space 
of vacua.  By supersymmetry, the moduli space must be hyper-\kahler for both 
the hypermultiplets and the (dualized) vector multiplets .  
Because of 
their different patterns of global R-symmetry breaking, however, 
the VEV's of the vector and 
hyper multiplets respectively are distinct order parameters of the 
theory, even after taking into account of quantum fluctuations.  
Vacua of N=4, D=3 gauge theories always contain a 
vector multiplet branch in which the gauge group is generically 
broken down to $U(1)^N$ 
where $N$ is the total rank of the gauge group.  This branch is 
parameterized by the $4N$ real scalars 
from the $N$ corresponding vector multiplets.  If it has a sufficient 
number of hypermultiplets, there can also be a hypermultiplet branch 
and/or mixed branches.
\btab {r-charge} {R charges of the VEV's and parameters} {llc}
Multiplets/Parameters & Notation & $SU(2)_{345} \times SU(2)_{789}$ \\ \hline
Vector &	$\vec a: a^3, a^4, a^5$ &	(3,1) \\ \cline {2-3}
&		$A_\mu \to \sigma$ &		(1,1) \\ \hline
Hyper &		$\vec r: r^7, r^8, r^9$ &	(1,3) \\ \cline {2-3}
&		$b$ & 			(1,1) \\ \hline
\FI &		$\vec \zeta: \zeta^7, \zeta^8, \zeta^9$& (1,3) \\ \hline
mass &		$\vec m: m^3, m^4, m^5$ &	(3,1) \\ \hline 
(electric) coupling & $e$ 		&	(1,1) \\
\etab

The low energy effective action up to two derivatives and four fermions 
is controlled by the metric of the moduli space, which depends on the 
parameters of the theory as well as the position on the moduli space.  
The dependence is constrained by extended supersymmetry: the \kahler potential 
is the sum of a term that depends only on the hypermultiplet scalar 
and one only on vector multiplet scalars \cite {wlp,aps:em}.  
\index{non-renormalization theorem}
So a mixed branch is 
the direct product of a vector branch $\Mv$ and a ``hyper'' 
branch $\Mh$.  By reinterpreting the parameters of the theory as the VEV's 
of background superfields, one can further deduce 
the effects of tuning them on the metric.  
The gauge coupling $\frac 1 {g^2}$ lives in a vector 
multiplet, so it can continuously deform the metric of $\Mv$ but not 
that of $\Mh$.  Since $g^2$ also plays the role of $\hbar$, 
one concludes that 
the the metric of $\Mh$ is determined purely 
classically whereas that of $\Mv$ in general 
receives quantum corrections.  Mass parameters also live in a vector 
multiplet, so they also can continuously 
deform $\Mv$ but not $\Mh$ - they can only 
affect $\Mh$'s dimensionality by changing the number of massless 
hypermultiplets.  
\FI parameters, on the 
other hand, live in hypermultiplet, and therefore can only deform $\Mh$ 
and change the dimensions of $\Mv$.  This is summarized 
in \tabr {parameter}.
\btab {parameter} {Parameters' influence on moduli space}
	{l  c  c  c  c}	
Branch &	$\vec m$& $\vec \zeta$ &$e$ &m (?)	 \\ \hline
$\Mv$ &		deform	& reduce & deform & no effect (?) \\ \hline
$\Mh$ &		reduce & deform & no effect & deform (?)\\ 
\etab

	Looking at the above two tables, the pattern for a possible 
duality emerges.  Starting with some theory that we call A model, 
if we switch what what we mean by  
$SU(2)_{345}$ and $SU(2)_{789}$, and at the same time exchange 
masses with \FI parameters, will we end up with 
an apparently different theory, model B, that is nonetheless 
equivalent to A?  $\Mv$ of one theory would have to be 
mapped to $\Mh$ of the other.  Classically this is definitely 
not true: $\Mh$ is a complicated space obtained via the  
hyper-\kahler quotient construction \cite {hklr} while $\Mv$ 
is just a flat space quotiented by the Weyl group for the gauge 
symmetry.  
Furthermore, when some of the mass terms $\vm_A$
of, say, A, vanish, enhanced global (flavor) symmetries emerge and act 
on $\Mh^A$.  For B there would have to be corresponding 
global symmetries emerging at vanishing \FI 
parameters $\vzeta_B$ and acting on $\Mv^B$.  
Classically there is no such symmetry.  Therefore this hypothetical 
duality, named \emph {mirror symmetry in three dimensions} 
\index{mirror symmetry} 
in \cite {is}, must be 
a quantum equivalence.  However, a glance at tables \ref {tab:r-charge}
and \ref{tab:parameter} 
reveals a missing dual of the gauge coupling constant $e$, whose property 
as predicted by mirror symmetry is listed in \tabr {parameter} 
with question marks.    
Brane interpretations of this parameter, known as the 
\emph {magnetic coupling} $m$ (without an arrow), \index{magnetic coupling}
have been proposed \cite {hw,hov}, but 
it has yet to be found in any Lagrangian formulation.  
Given such, mirror symmetry can only manifest for ordinary 
field theory when $e$ approach a particular value\footnote 
	{However, it is possible, and the brane realization discussed 
	later strongly suggests, that $m$ can be a field theory parameter 
	that has no Lagrangian representation.}.
Being dimensionful 
($e^2$ has the dimension of mass in 3d), the only natural candidates are 
$0$ and $\infty$.  $e=0$, the classical limit, is already ruled 
out.  So we are left with the strong coupling limit, which, since it 
is in 3d, is also the infrared limit.  This also leads us to one of 
the most striking aspects 
of this proposed mirror symmetry: it maps from one theory 
the metric for $\Mv$, a quantity that 
receives very large quantum corrections, 
to, in the dual theory, the metric for 
$\Mh$, which is given by purely classical expressions.  
As many physicists have suggested, this may imply  
the line between quantum and classical physics is more blurred  
than previously thought.

Because of its strong coupling nature, proving mirror symmetry 
within the context of field theory will be difficult.  
Embedding the field theory in the dynamics of branes 
\cite 
{douglas:gauge-field,sen:f-theory,bds:probe,dkps:short-distance,seiberg:three-d} 
renders many aspects of mirror symmetry 
manifest \cite {hw,bhooy}, if 
one assumes the S duality of type IIB string.  
This will be reviewed extensively in the 
rest of this talk.  Before that, I want to give some example of 
mirror pairs and one of 
the many pieces of field theoretic evidence 
supporting it \cite {bhoo}.  
They are logically independent of any conjectures about string theory.

	A model has gauge group $U(K)$ with $N$ fundamentals and 
1 adjoint hypermultiplets.  B models has gauge group $U(K)^N$,
which we label as $U(K)_\alpha$, 
$\alpha = 0, \ldots, N-1$. Its hypermultiplets consist of one fundamental 
charged under $U(K)_1$ and N bi-fundamentals.  The latters 
are each charged respectively under 
$U(K)_{\alpha} \times U(K)_{\alpha +1}$ in representation $(\bar K, K)$, 
with cyclic identification 
$\alpha \sim \alpha + N$.  These field contents are nicely 
encoded in the ``quiver'' diagrams%\footnote
%	{They are formerly known to physicists as ``moose diagrams'' 
%	\cite {susskind:m,georgi:m}.} 
\cite {dm:ale} \index{quiver diagram}
of figure 1.  Each inner node with a number $K$ represents a 
$U(K)$ gauge group.  Each link connecting a pair of them represents 
a bi-fundamental charged under the pair as 
(fundamental, fundamental*).  An outer node with number $N$ 
represents fundamentals with multiplicity N charged under the 
gauge group associated with the inner node to which it is attached.

\putfig {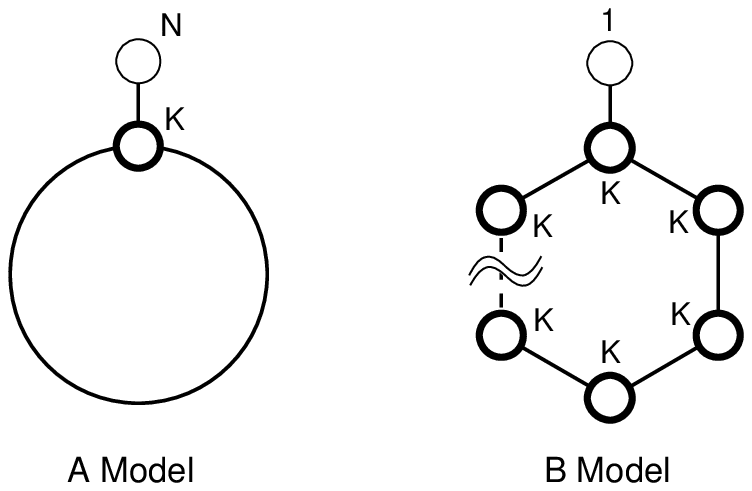} {\FigSizeSimpleQuiver} {Quiver for A and B models.}

Using the notation given in table 1, the metric for $\Mv^A$  
takes the form
\beq	\label {eq:Mv-metric}
ds^2 = g_{ij}d\va_i d\va_j + (g^{-1})_{ij}
	(d\sigma_i + \vomega_{il}\cdot d\va_l)
	(d\sigma_j + \vomega_{jl}\cdot d\va_l),
\eeq
under the constraint 
\beqar
{\vgrad}_i g_{jl} & = & \vgrad_j g_{il} \\
\dod {a_i^x} \omega^y_{jl} - \dod {a_j^y} \omega^x_{il} 
& = & \epsilon_{xyz} \dod {a_i^z} g_{jl}.
\eeqar
Here $i, \dots = 0,\dots, k-1$ index the Cartan of $U(k)$; 
$x, \ldots = 3, 4, 5$.
In \cite {bhoo}, $g_{ij}$ is computed.  Perturbatively it 
is one-loop exact: 
\beqar	\label {eq:one-loop}
	g_{i\neq j} &=& \frac{2}{|\va_i-\va_j|}
	- \frac{1}{|\va_i-\va_j +\vm_{\adj}|}
	- \frac{1}{|\va_i-\va_j -\vm_{\adj}|}, \nono
g_{ii} &=& \frac{1}{e^2}
	+ \sum_{\alpha=0}^{N-1}\frac{1}{|\va_i-\vm_\alpha|} \\
	& & + \sum_{j=1\dots K}^{j\neq i}
		\left( \frac{-2}{|\va_i-\va_j|}
	+ \frac{1}{|\va_i-\va_j +\vm_{\adj}|}
	+ \frac{1}{|\va_i-\va_j -\vm_{\adj}|} \right). \nono
\eeqar
Here $\vm_\adj$ is the triplet mass for the adjoint hypermultiplet; 
$\vm_\alpha$ those of the fundamentals, indexed by 
$\alpha = 1,\ldots,N$.  
When $\vm_\adj = 0$, there is no instanton correction to the metric and 
\eqr {one-loop} is also nonperturbatively exact.  For illustration, 
consider this simpler case, so that $g_{i\neq j}$ 
vanishes. \eqr {Mv-metric} and \eqr {one-loop} state 
that $\Mv^A$ is the direct product 
of $K$ multi-Taub-NUT space with charge $N$.  \index{Taub-NUT space}
After quotiented by 
the Weyl group of $U(K)$, the direct product becomes a symmetric product. 

	It turns out that setting $\vm_\adj = 0$ for A model is mapped 
by mirror symmetry to a condition on the \FI parameters $\vzeta_\alpha$
of B model:
\[	\sum_\alpha \vzeta_\alpha = 0.	\]
In this case, $\Mh^B$ 
is given by the symmetric product of $K$ ALE spaces with $A_{N-1}$ 
singularity \cite {kn}.   \index{ALE space} 
The metric of each ALE space is given by 
\[	\label{eq:selfdual}
ds^2 = g d\vr^2 + \frac 1 g (db + \vomega (\vr) \cdot d\vr )^2
\]
with 
\[	\label {eq:ale}
	g (\vr) = \sum_{\alpha = 0}^{N-1} 
		\frac 1 {\abs {\vr - 
			\sum_{\beta = 0}^{\alpha} \vzeta_\beta}}.
\]
Now the metric for each Taub-NUT factor of $\Mv^A$ 
can be read from \eqr {one-loop} after setting $\vm_\adj = 0$.  
\[	\label {eq:Taub-NUT}
	g (\va)= \frac 1 {e^2} + \sum_{\alpha = 0}^{N-1} 
		\frac 1 {\abs {\va - \vm_\alpha}}.
\]
It is clear that $\Mv^A = \Mh^B$ if and only if 
$e = \infty $ and one makes the following 
identification\footnote
	{To see the counting of parameters matches, note 
	that the ``center'' of the mass parameters can be 
	absorbed by shifting the origin of $\Mv$ on 
	both sides.  Here this is used to set 
	$\vm_{N-1} = 0$ for A model.}
\[	\label {eq:vm-zeta-mirror}
	\vzeta_\alpha = \vm_\alpha - \vm_{\alpha -1}.
\]

\section {Set the Branes to work}

	Consider in type IIB string theory, a configuration 
that includes 3 types of branes, whose 
worldvolume configurations 
are given in \tabr {brane-config} \cite {hw}.  
\btab {brane-config} {Configurations of the branes} 
	{ l  c  c  c  c  c  c  c  c  c  c } 
 & 0 & 1 & 2 & 3 & 4 & 5 & 6 & 7 & 8 & 9 \\ \hline
D3-brane & \x & \x & \x & & & & \x & & & \\ \hline
D5-brane & \x & \x & \x & & & & & \x & \x & \x \\ \hline
NS5-brane & \x & \x & \x & \x & \x & \x & & & & \\ 
\etab
Such a configuration can preserve up to 8 supercharges, corresponding 
to N=4 in 3d.  Of the original $Spin(1,9)$ Lorentz symmetry, only 
the (1+2)d Lorentz group $SL(2,R)_{012}$ and the R-symmetry group 
$SU(2)_{345} \times SU(2)_{789}$ remain manifest.  
I will now review some basic rules for ``model building'' 
from branes and their justifications.  Many of them first 
appeared in \cite {hw}.  \index{brane engineering}

\renewcommand {\labelitemi} {$\bullet$}
\begin {itemize}

\item 	By sending $\Mp$ to $\infty$, the worldvolume theory on the branes 
decouple from the  bulk fields such as graviton.  \index{brane probe}

\item	D3-branes can break and end on D5 or NS5-branes without violating 
RR charge conservation \cite {strominger:open-p,townsend:d-from-m} by becoming a 
magnetic ``monopole'' on the 5-branes, as depicted 
on the left in 
\figr {BraneBreaking}.  There the horizontal lines represent D3-branes; 
solid and dashed vertical lines represent NS and D5-branes respectively.  
Conversely, two D3-branes ending on the same 5-brane from opposite sides 
can rejoin.
\putfig {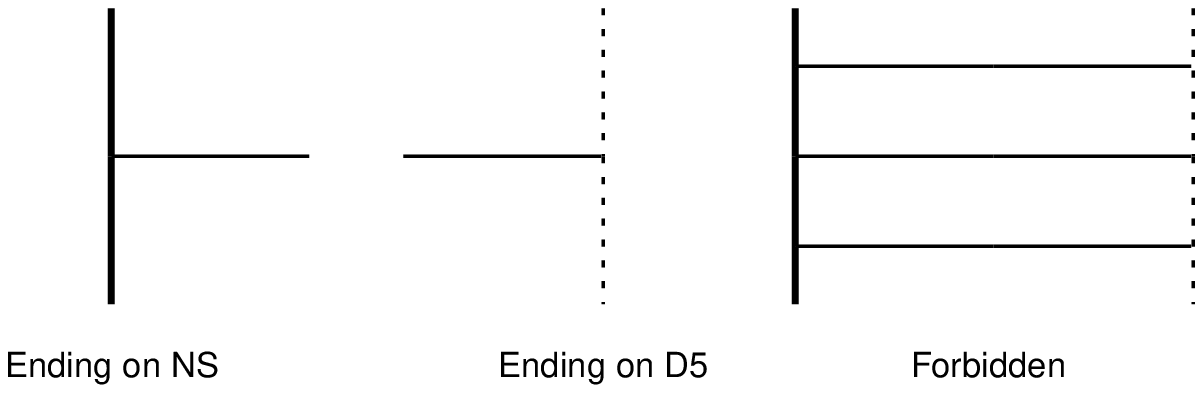} {\FigSizeBraneBreaking} 
	{D3-brane ending on 5-branes and forbidden configurations.}

\item	To the worldvolume theory on the D3-branes, breaking and ending 
are tantamount to 
imposing boundary conditions, reducing N=4, D=4 supermultiplets to 
N=4, D=3 supermultiplets.

\item	By taking the appropriate scaling limit, the Kaluza-Klein modes 
along the 6th direction can be kept massive and integrated out.  
The effective infrared theory on the D3-brane is therefore a (1+2)d QFT.

\item	The worldvolume theories on the 5-branes are weakly coupled in the 
infrared.   The fields on them have an infinite volume coefficient in their 
kinetic terms as compared to D3-brane fields due to their relative 
sizes.  As a result they are 
frozen as background fields and their VEV's are parameters of the 
effective 3d 
theory.  Gauge symmetries on the 5-branes become global symmetries.

\item	When a NS5-brane crosses a D5-brane, a D3-brane is created and 
stretched in between.  This is a consequence of charge conservation 
\cite {hw}\footnote
	{For some other treatments of this phenomenon, 
	see \cite {klebanov:review} 
	and the references therein.}. 

\item	Certain configurations are believed to be forbidden.  They 
involve more than one D3 brane stretched between the same NS-D5 pair, 
such as the one in the right of \figr {BraneBreaking}.

\end {itemize}

Following the above rules one can build brane configurations of arbitrary 
complexity.  To read off the contents of the resulting field theory, 
one also needs to know the following.

\begin {itemize}

\item	Dynamical fields of the decoupled (1+2)d theory arise out of the lightest 
excitations of open strings starting 
and ending on the D-branes.  There are three types: open strings connecting 
between D3-branes in the same ``cubicle'' (\figr {ElecField}a) 
give vector multiplets; those between D3-branes in adjacent 
``cubicle'' give bi-fundamental hypermultiplets (\figr {ElecField}b); 
while open strings connecting 
between D3 and D5 give fundamental 
hypermultiplets (\figr {ElecField}c).  These can be read off from  
perturbative open string quantization and the boundary conditions.  Their 
transformation properties under R-symmetry agree with the assignment 
given in the last section.  Their end points are electric sources 
on the D3-branes.
\putfig {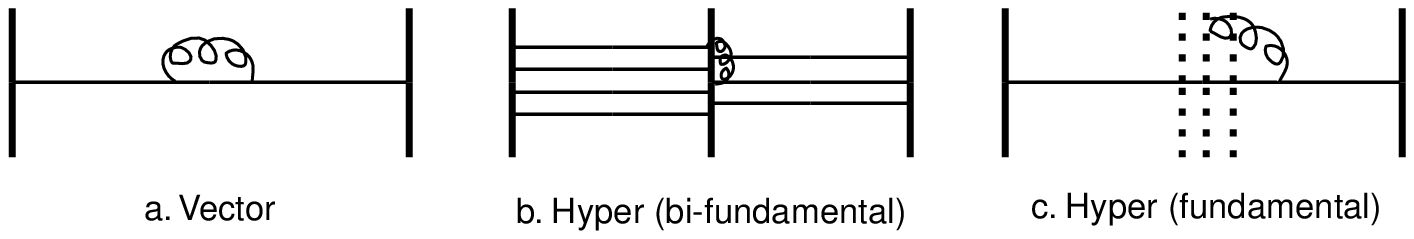} {\FigSizeElecField} 
	{Field content from open fundamental strings.}

\item	Configuration of the D3-branes selects the gauge symmetries
and the vacuum, as illustrated in \figr {DThreeConfig}.
\putfig {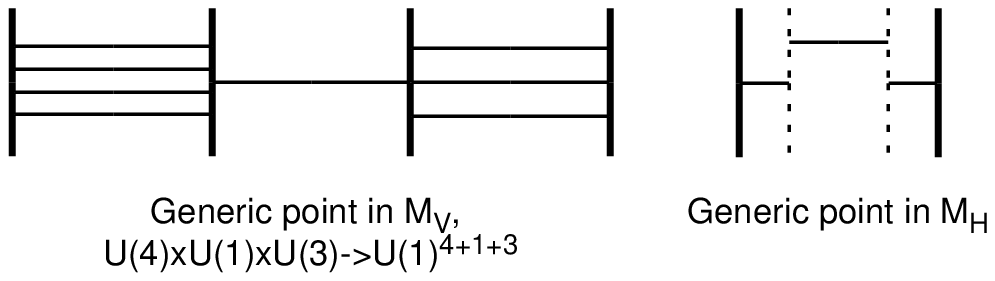} {\FigSizeDThreeConfig} {D3-brane configuration selects gauge 
	group and vacuum}

\item	Configuration of the 5-branes determines the parameters 
(\figr {FiveBraneConfig}) and global 
symmetries of the 3d QFT.  Note the identification 
of the magnetic coupling constant as the inverse square root of the 
distance between adjacent D5-branes along the 6th direction.  The frozen 
gauge dynamics of the 
D5-brane gives rise to the flavor global symmetry acting on the fundamental 
hypermultiplets.  Since in conventional Lagrangian field theories, 
this global symmetry is restored by setting the masses to zero, 
the magnetic coupling is fixed to be infinite.  At the same time, 
the global symmetry resulting from the NS5-branes is 
broken unless they coincide, which amounts to setting \FI to zero 
and $e$ to infinity --- it can only appear at a nontrivial infrared 
fixed point.
\putfig {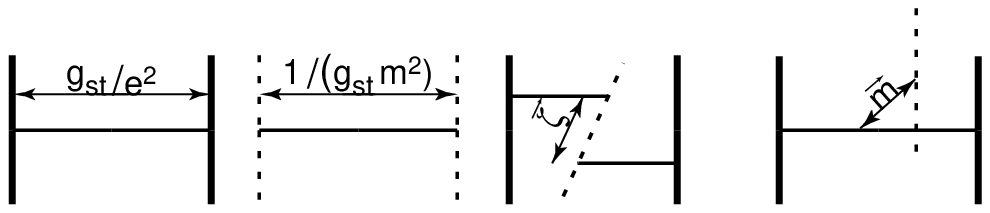} {\FigSizeFiveBraneConfig}
	{5-brane configuration fix parameters and 
	global symmetries}

\item	Type IIB string theory has a nonperturbative \emph {S} duality 
\cite {schwarz:sl-two-z} \index{S duality}
that inverts the string coupling constant and exchange NS5-branes 
with D5-branes as well as fundamental strings with D-strings.
It leaves invariant D3-branes but acts on their worldvolume theories 
as the S duality for N=4, D=4 SYM \cite {gg}.

\item	We therefore should also consider excitations of open 
D-strings starting and ending on D3-branes and/or NS5-branes.  
They are simply obtained from the S dual of \figr {ElecField}.  
However, on D3-branes, rather than generating additional degree 
of freedom, they are related to   
the open string fields nonlocally --- end points of these two types 
of string on D3-branes 
are respectively magnetic and electric sources.  Each appear 
as solitonic excitations of the other and are exchanged by 
the field theoretic S duality.

\item	Therefore there are two descriptions of the same 
3d theory: one uses open fundamental string fields as the 
canonical variables 
while the other uses open D-string fields.  Their equivalence 
is the 3d mirror \index{mirror symmetry! and S duality}
symmetry, and it follows from the S duality type IIB string theory.

\item	It is convenient to rephrase this duality as an operation 
combing an S duality transformation with the interchange of $345$ and 
$789$ directions \cite {hw}.  This makes explicit the interchange of 
the two R-symmetry factors, $\Mv$ with $\Mh$, 
masses with \FI parameters, and $e$ with $m$.  For reasons just stated, 
this is reflected on ordinary Lagrangian field theories 
only in their infrared limit.

\end {itemize}

\section {Mirror Pairs}

	Now I will present a few examples of mirror pairs constructed 
in \cite {bhooy}.  Their field 
contents are best described by the type of 
quiver diagrams introduced earlier.  
They are obtained using the brane-engineering rules outlined above, but 
with a compactified 6th direction, i.e. with periodic identification 
along $X^6$.
As a warm-up, let's look at the A and B models given in figure 1.  
The corresponding brane configurations are sketched in \figr {SimpleBrane}
\putfig {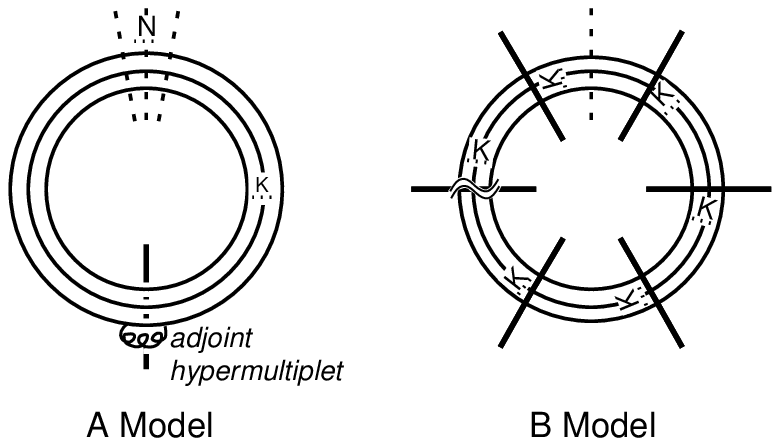} {\FigSizeSimpleBrane}
	{Brane configurations for the quivers in \figr {SimpleQuiver}.}

For A model, the $N$ fundamentals of $U(K)$  originate from open strings 
stretched between the $N$ D5 and the $K$ D3-branes.  A special feature 
of this configuration is that the bi-fundamental, coming 
from open string 
stretching between ``nearest-neighbor'' D3-branes, become the adjoint of 
$U(N)$ because there is only one gauge groups.  For B model, 
more generic situation prevails and there are $N$ bi-fundamentals.  
The correspondence between the moduli spaces of vacua of A and B 
as well as the identification \eqr {vm-zeta-mirror} is evident.  
As it is, there is one constraint on the field theory parameters, namely 
\[	\vm_\adj = 0 = \sum_\alpha \vzeta_\alpha.	\]
A way to relax this condition has been given in \cite {w:four-d}.
\index{mirror symmetry!examples}	

	Now let's look at mirror pairs of more complex theories.  
As depicted by the quiver diagrams in 
\figr {UkGeneralFlavor}, model A again has 
gauge group $U(K)^N$ and $N$ bi-fundamentals, but each $U(K)$ now 
has fundamentals with an arbitrary number of flavors $w_i$.  Its mirror, model 
B, has gauge group $U(K)^M$  with 
\[	M = \sum_i w_i.		\]
Besides the 
$M$ bi-fundamentals, it has $N$ fundamentals arranged as shown in 
\figr {UkGeneralFlavor}.  If all $w_i > 0$, 
each $U(K)$ factor has at most one fundamental. 
If some $w_i = 0$, the corresponding nodes 
in B model's quiver 
coalesce and give rise to fundamentals of higher flavor.  
Such mirror pairs are again constructed via the S duality of 
type IIB string theory \cite {bhooy}.  The mirror map relates 
the moduli spaces of the two theories, as well as their parameters, 
in the same manner as the simpler case discussed above.
\putfig {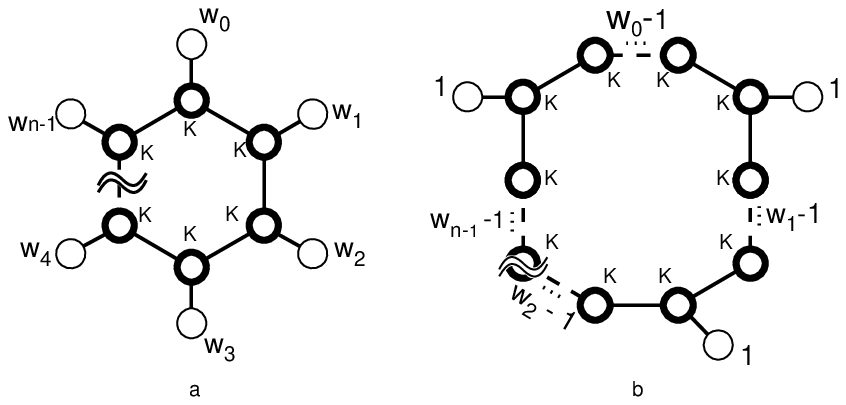} {\FigSizeUkGeneralFlavor} 
	{Arbitrary flavor of fundamentals and the mirror}

	It is natural to generalize to the cases with A model 
having gauge group 
$\prod_{i= 0}^{N-1} U(K_i)$, $N$ bi-fundamentals, and arbitrary 
fundamentals.  Its quiver is shown in \figr {GeneralCase}, 
along with its brane realization.  
\putfig {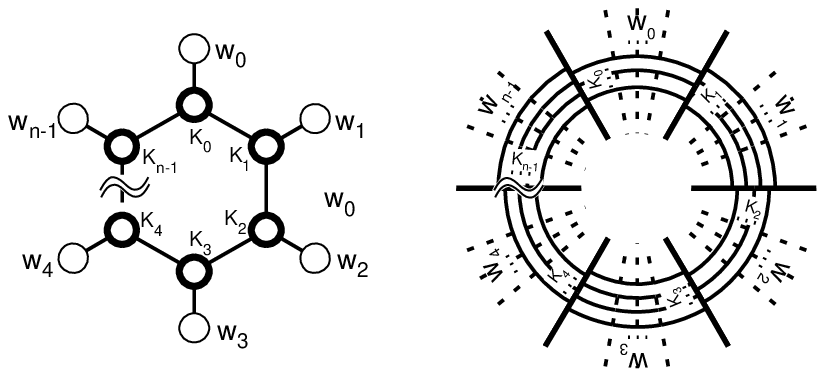} {\FigSizeGeneralCase} {Quiver and brane realization 
	of $\prod_i U(K_i)$}
However, here one encounters an important subtlety. 
Although one can always perform a S duality transformation and obtain 
the mirror configuration, the result does not always correspond 
to a gauge theory.  To see this, recall that mirror symmetry exchanges 
$\Mv$ with $\Mh$.  A universal property of N=4, D=3 super-Yang-Mills 
theories is the existence of the Coulomb phase, a branch of $\Mv^B$ with $4r$ 
dimensions, where $r$ is the total rank of the gauge group.  This is 
mapped under mirror symmetry to a branch of $\Mh^A$: the completely Higgsed 
phase.  Therefore a necessary condition for model B to have an ordinary 
gauge theoretic Lagrangian description is for model A to admit complete 
Higgsing.  This amounts to requiring  
\cite {nakajima:ins-ale-quiver-km,nakajima:quiver-km,bhoo,bhooy}:
\beq	\label {eq:complete-higgsing}
	2k_i - k_{i-1} - k_{i+1} \leq w_i.
\eeq
When this is satisfied, the mirror gauge theory can be 
constructed along the same vein as before.  The details become  
complicated and can be found in \cite {bhooy}.  

\section {Phases and Transitions}

	What happens if \eqr {complete-higgsing} is not satisfied?  
S duality still gives a mirror configuration, but one without 
a gauge theoretic description.  An example of this is shown in 
\figr {NonLagrangian}a.
\putfig {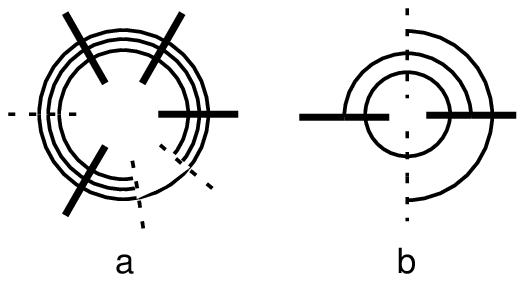} {\FigSizeNonLagrangian}
	{Brane configurations giving rise to novel field theories} 
	
To understand this phenomenon, note 
that from the field theory perspective, mirror symmetry corresponds 
to the $Z_2$-wise freedom in labeling the two $SU(2)$ R-symmetries.  
While the vector and hypermultiplets transform distinctly though 
somewhat symmetrically under them, their interactions 
enter in the Lagrangian in rather different form.  The Lagrangian 
descriptions of a theory and its dual would in general be quite different, as 
the examples above show.  Actually, there is no reason a priori 
to expect that both sides of a mirror pair have Lagrangian 
descriptions at all.  It is natural to conjecture 
this is the what is happening 
here\footnote 
	{In fact, barring an unexpected way to incorporate the mysterious 
magnetic coupling into a Lagrangian formulation, even for the cases in which 
both of the dual pair have a Lagrangian description, 
mirror symmetry is manifest 
only at the infinite coupling, i.e. infrared limit, as mentioned earlier.  
Here I shall be referring to that limit implicitly unless otherwise stated.}.
3d N=4 theories with Lagrangian description 
can therefore be classified into those that 
have two, 
related by mirror symmetry in the infrared, and 
those that have only one\footnote 
	{One should note that the lack of a completely higgsed phase 
	is necessary but not sufficient for a non-Lagrangian dual.  
	For example, a free $U(1)$ vector multiplet, which obviously 
	does not have any Higgs phase, is dual to a free hypermultiplet.  
	However, such free theories 
	cannot help explain interacting fixed points, e.g. 
	when the D3-branes in \figr {NonLagrangian}a or b coincide.}.
	
An example of the latter type, in which the non-Lagrangian description 
is that with $E_n$ tensionless string \cite {gh:e-eight,sw:six-d}, 
was conjectured already in \index{tensionless string}
\cite {is}.  It has recently been explicitly 
engineered using one of the alternative 
formulations of mirror symmetry from string theory 
\cite {hov,kmv:mirror-four-d}.  Here we 
have a very simple prescription for engineering an infinite number of 
such Lagrangian-non-Lagrangian mirror pairs.  As noted 
in \cite {is}, these are local quantum field theories, simply because 
on one side of the mirror there is a Lagrangian description that flows 
to it.
However, experience in 2, 4, 5, and 6 dimensions 
has shown 
a Lagrangian description, though convenient in many ways, 
may not be a necessary condition for a local quantum 
field theory (see, for example, 
\cite {ad:em,seiberg:five-d,seiberg:six-d,seiberg:sixteen}.  
\index{non-trivial fixed point} 
\index{mirror symmetry!non-trivial fixed point}
Indeed, one can easily engineer using branes 
a third class of theories that have 
\emph {no} Lagrangian description on either side of the mirror.  
An example is 
shown in \figr {NonLagrangian}b.  Since the decoupling of bulk 
as well as Kaluza-Klein modes works just as in the more mundane 
cases, they should still be interacting 
local quantum field theories, but 
with no known Lagrangian description flowing to it.

	Such interesting  
phenomena deserve an explanation 
from string theory. In that context mirror symmetry 
is the equivalence of two descriptions of the same physics related 
by S duality.  The degrees of freedom on the D3-brane theory can be 
captured both by open fundamental string and by open D-string 
fields.  Starting with, say, a description using only open fundamental 
string fields, one 
can employ standard string perturbation theory to obtain their 
interactions and write a Lagrangian for the 
field theory modes in the decoupling limit.  
The same 
prescription goes through for a description based solely on open D-string 
fields.   Suppose, however, that there is no way to capture the full 
degrees of freedom 
by using only, say, open D-string fields.  
The description on the B model  
side may not be in the form of a local action, as 
open D-strings and open fundamental strings ending on D3-branes are 
mutually nonlocal.  If neither open fundamental string fields nor open 
D-string fields can account for the full degrees of freedom by themselves 
respectively, we end up with the third class of theories.

\putfig {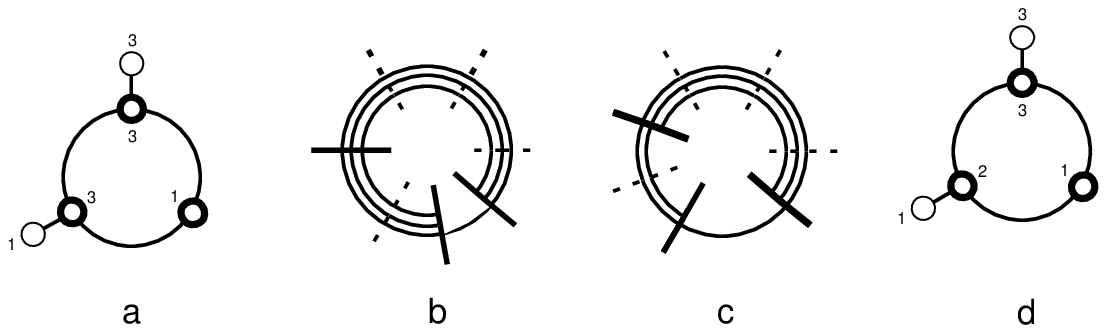} {\FigSizeTransition} 
	{Transition between different classes of field theories 
		via brane motion.}
	Remarkably, with branes one can not only engineer 
examples of all three types 
of theories, but also interpolate between them continuously 
by moving the 5-branes around.  \index{phase transition} 
Shown in \figr {Transition}a, b is the mirror of \figr {NonLagrangian}a.  
It is a theory of the 
second type (Lagrangian-non-Lagrangian mirror pair).  By moving 
one NS5-branes past another, we arrive at the theory depicted in 
\figr {Transition}c, d, which is of the first type 
(Lagrangian-Lagrangian mirror pair).  Similar transition can be engineered 
between theories of the third class and the first two as well.
To appreciate the meaning of such a process, recall that the 
electric and magnetic coupling constants 
are inversely proportional to the 
square root of the distance between adjacent NS5 and adjacent 
D5-branes respectively.  Moving 5-branes of the same type past each 
other effectively 
makes coupling constants of the corresponding type 
imaginary.  This is indicative of a change of the effective 
degrees of freedom describing the system, namely a phase transition.  That 
smooth movements in the moduli space of brane configurations can connect 
distinct classes of field theories via some type of phase transitions 
is one of the most important lessons to be learned from this work.

\acknowledgements
I would like to thank the organizers of this summer institute  
for invitation, financial support, and the opportunity to present this 
talk, as well as for making the whole school a rewarding experience.  
I am also indebted to J.~de~Boer, K.~Hori, H.~Ooguri, and Y.~Oz 
for collaboration and many useful discussions.  I would like to especially 
acknowledge H.~Ooguri for encouragement.  My research is supported 
in part by DOEAC03-76SF00098 and a grant from the Dean of Graduate Division, 
U. C. Berkeley.

\end{document}